\documentclass[preprint]{aastex}
\newcommand{\dif}{\mathrm{d}}
\renewcommand{\vec}[1]{\mbox{\boldmath$#1$}}
\slugcomment{submitted to PASP}
\shorttitle{Polarization Confidence Limits}
\shortauthors{Vaillancourt}
\begin{document}

\title{Placing Confidence Limits on Polarization Measurements}
\author{John E. Vaillancourt}
\affil{Enrico Fermi Institute, University of Chicago, 5640 S. Ellis Ave., Chicago, IL 60637}
\email{johnv@oddjob.uchicago.edu}

%\maketitle

\begin{abstract}
The determination of the true source polarization given a set of
measurements is complicated by the requirement that the polarization
always be positive. This positive bias also hinders construction of
upper limits, uncertainties, and confidence regions, especially at low
signal-to-noise levels. We generate the likelihood function for linear
polarization measurements and use it to create confidence regions and
upper limits.  This is accomplished by integrating the likelihood
function over the true polarization (parameter space), rather than the
measured polarization (data space).  These regions are valid for both
low and high signal-to-noise measurements.
\end{abstract}
\keywords{
%  methods: statistical ---
  polarization
}

\section{Introduction}

Measurements of linear polarization generally provide values for the
degree and angle of polarization.  They may also provide upper limits
on the degree of polarization even where the nominal values $P$ are
not much larger than the uncertainties $\sigma$ (i.e.\ at low
signal-to-noise).  Polarization maps typically plot vectors in regions
where measurements are made with high confidence and open circles
%are often plotted in regions 
where the polarization is found to be too low to justify a vector, but
where useful upper limits can still be estimated (e.g.\ where
$P+2\sigma<1\%$; \citealt{archive}).  

While the distributions of the polarization degree and angle are not
normal (gaussian), it is still possible to place estimates and
uncertainties and confidence levels on their values.
%, (3) a best estimate of the polarization position angle, and (4)
%uncertainties and confidence levels on the position angle.
The position angle distribution, while non-normal, is symmetric about
the measured angle so that estimates and uncertainties on its value
are fairly straightforward \citep{clarke93}.  Estimates of the
polarization degree are complicated by the fact that it must always be
positive, introducing a bias into any estimate. For any true
polarization degree $P_0$, we expect on average to measure a
polarization degree $P>P_0$. This problem has been thoroughly
investigated by \citeauthor{simmons85} (\citeyear{simmons85};
hereafter SS85).

SS85 have also devised a procedure to place uncertainties on the
polarization degree.  First, the expected probability density of
polarization (\S\ref{sec-distrib}) is integrated over a set of
possible measurements.  However, this only places limits on the
measured polarization $P$, the limits varying with the value of the
true polarization $P_0$.  To correct for this and thus create
confidence limits on the parameter $P_0$, SS85 project their results
onto the $P_0$-axis.

Here we introduce a slightly different procedure for determining
confidence regions on the true polarization.  Rather than integrate
the probability density function over the data-space of the measured
polarization, we integrate over the parameter-space of the true
polarization.  In this way we directly determine the likelihood
(\S\ref{sec-likelihood}) that a measured polarization $P$ was drawn
from a source of true polarization $P_0$.

\section{Rice Distribution} \label{sec-distrib}

Measuring the degree and angle of polarization is equivalent to
determining of the amplitude and phase of a sinusoidal signal in the
presence of noise.  The joint and marginal probability distribution
functions for the amplitude and phase have been discussed in detail by
other authors \citep{rice45,vinokur65,simmons85}.  For completeness we
review their results here.

Consider the simple case of rotating a linear polarizer through some
angle $\theta$.
%Polarization is measured by modulating incoming radiation with an
%optically active element.  In the simplest case, a signal can be
%measured as a linear polarizer is rotated through some angle $\theta$.
The resulting signal changes as
\begin{equation}
S(\theta) = P \cos 2(\theta + \phi) + S_N,
\label{eq-polsig}
\end{equation}
where the amplitude $P$ is the polarization degree, $\phi$ a relative
phase shift (the polarization angle), and $S_N$ gaussian random noise.
This form can be decomposed into two orthogonal components
\begin{equation}
S(\theta) = (q + q_N) \cos2\theta - (u + u_N) \sin2\theta,
\end{equation}
where 
\begin{eqnarray}
q & = & P\cos2\phi,\label{eq-q}\\
u & = & P\sin2\phi \label{eq-u},
\end{eqnarray}
$q_N$ is the noise component in phase with $\cos2\theta$ and $u_N$ is
the noise component in phase with $\sin2\theta$.  In optics $q$ and
$u$ are known as Stokes parameters.  In electronics they are often
referred to as the ``in-phase'' and ``quadrature-phase'' components,
respectively.\footnote{The optical polarization vector has no
  direction and is therefore periodic in $\pi$, as is clear from the
  factor of 2 in the arguments to the sine and cosine functions.  In
  electronics the signal is periodic in $2\pi$ by definition, so the
  factor of 2 is replaced by unity in equations (\ref{eq-polsig}) --
  (\ref{eq-u}).}

Since $S_N$ is gaussian, the measured Stokes parameters $q$ and $u$
are normally distributed about the true values $q_0$ and $u_0$ with
equal uncertainties $\sigma$ (i.e.\ $\langle q_N^2 \rangle = \langle
u_N^2 \rangle = \sigma$).  The values of $q_0$ and $u_0$ are given by
equations (\ref{eq-q}) and (\ref{eq-u}), replacing $P$ by $P_0$ and
$\phi$ by $\phi_0$.
The uncertainty in the polarization $P$ is also given by $\sigma$.
%For simplicity we define the signal-to-noise relations $p = P/\sigma$
%and $p_0 = P_0/\sigma$.  We will refer to the lower-case $p$'s as
%polarization degrees throughout the remainder of this work;
%however, they are truly dimensionless signal-to-noise ratios.

The joint probability distribution for $q$ and $u$ is given by the
product of their individual normal distributions.  Transforming to the
polar coordinates $P$ and $\phi$ yields
\begin{equation}
f(P,\phi) = \frac{1}{2\pi\sigma^2}\exp\left[-\frac{P^2 + P_0^2 - 2PP_0\cos2(\phi-\phi_0)}{2\sigma^2}\right].
%f(p,\phi) = \frac{1}{2\pi}\me^{-\onehalf[p^2 + p_0^2 - 2pp_0\cos2(\phi-\phi_0)]}.
\label{eq-joint}
\end{equation}
The probability of measuring a polarization in the range [$P$, $P+\dif
  P$] and [$\phi$, $\phi+\dif\phi$] is given by $f(P,\phi)\,P\,\dif
P\,\dif(\phi-\phi_0)$.  Integrating over $\dif(\phi-\phi_0)$ yields
the Rice distribution for the polarization degree:
\begin{eqnarray}
F(P|P_0) \, \dif P & = &
\frac{1}{2} P\, \dif P \int\limits_{-\pi/2}^{\pi/2} f(P,\phi) \, \dif(\phi-\phi_0) \nonumber \\
& = & \frac{P}{\sigma} \exp\left[-\frac{P^2 + P_0^2}{2\sigma^2}\right] I_0\left(\frac{PP_0}{\sigma^2}\right) \frac{\dif P}{\sigma},
%& = & p \,\me^{-\onehalf\left(p^2 + p_0^2\right)} I_0(pp_0) \,\dif p,
\label{eq-rice}
\end{eqnarray}
%\begin{equation}
%G(\phi-\phi_0,p_0) = \int_0^{\infty} f(p,\phi) \, \dif p = 
%\frac{\me^{-(p_0^2/2)}}{\sqrt{\pi}} \left\{ \frac{1}{\sqrt{\pi}} + \eta_0 \me^{\eta_0^2} \left[ 1+\erf (\eta_0) \right] \right\}
%\label{eq-phid}
%\end{equation}
where $I_0()$ is the zeroth order modified Bessel function. (The
pre-factor of $\onehalf$ in the integral of eq.\ [\ref{eq-rice}]
corrects for the $\pi$-periodicity of $\phi$.)
%erf() is the Gaussian error function, and $\eta_0 = (p_0/\sqrt{2}) \,
%\cos2(\phi-\phi_0)$.

\begin{figure}
\epsscale{0.45}
\plotone{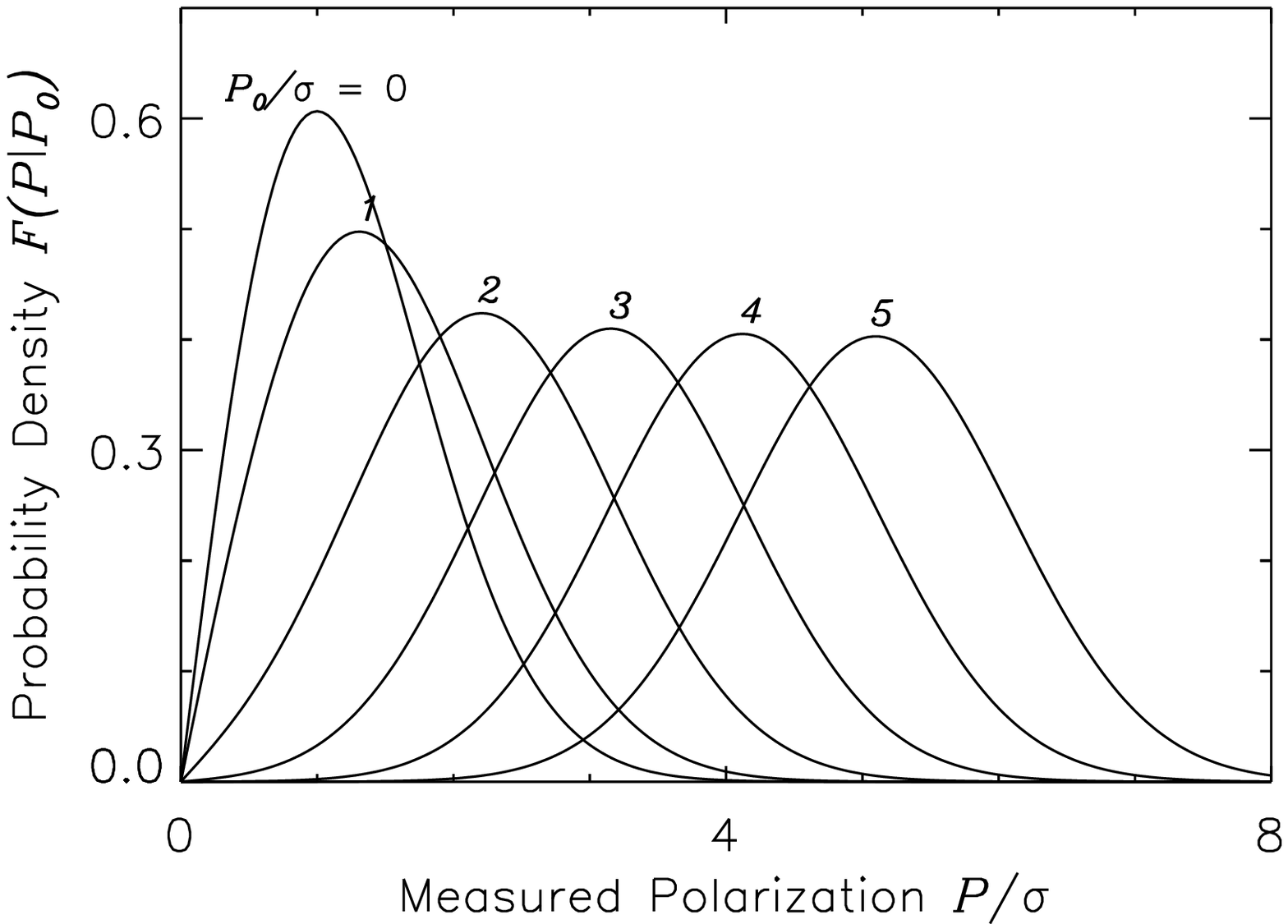}
%\plotone{ricedistrib.eps}
\caption{Probability density distribution for the measured
  polarization $P/\sigma$, given by equation (\ref{eq-rice}). Distributions
  are shown for several different values of the true polarization
  $P_0/\sigma$. \label{fig-rice}}
\end{figure}

The asymmetry of this distribution with respect to $P$ and $P_0$
(evident in Fig.\ \ref{fig-rice} for low values of $P_0/\sigma$) results in
the positive bias of the measured polarization.  For high
signal-to-noise the Rice distribution is near-normal with mean
approaching $P_0$ and a standard deviation approaching $\sigma$ (e.g.\ \citealt{rice45}).
%\begin{equation}
%%\lim_{pp_0\rightarrow\infty}
%F(p,p_0) \approx
%\left(1+\frac{1}{8pp_0}\right)
%\left(\frac{p}{2\pi p_0}\right)^{1/2} \exp\left[-\frac{(p-p_0)^2}{2}\right]
%%F(p,p_0) \approx
%%\left(1+\frac{1}{8pp_0}\right)
%%\sqrt{\frac{p}{2\pi p_0}} \, \me^{-\frac{1}{2}(p-p_0)^2}
%\label{eq-riceinf}
%\end{equation}
% ----- FIGURE 1 -----

\section{The Likelihood Function} \label{sec-likelihood}

\subsection{Constructing the Polarization Likelihood}

\begin{figure}
\plotone{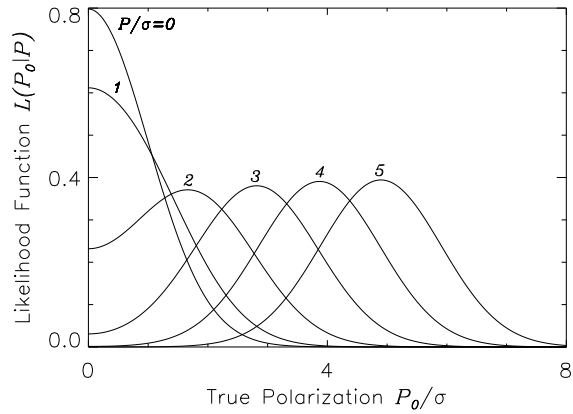}
%\plotone{likelihood.eps}
\caption{Normalized likelihood function $L(P_0|P)$ given by equation
  (\ref{eq-likeli}). Distributions are shown for
  several different values of the measured polarization $P/\sigma$.
  \label{fig-rice2}}
\end{figure}

If the true polarization $P_0$ is known, the probability that a
measurement will be made in the range [$P$, $P+\dif P$], is given by
equation (\ref{eq-rice}). However, one is more likely to have
knowledge of the measurement $P$ and wish to know the probability that
the true polarization is within the range [$P_0$, $P_0 + \dif P_0$].
We require the likelihood function $L(P_0|P)$ such that the aforementioned
probability is given by $L(P_0|P)\,\dif P_0$.  For a model
characterized by parameters $\vec{y} = \{y_1, y_2, \ldots, y_m\}$, and
a measured dataset $\vec{x} = \{x_1, x_2, \ldots, x_n\}$, the
likelihood function is constructed from
\begin{equation}
L(\vec{y}|\vec{x}) = \prod_{i=1}^n \; f(x_i \vert \vec{y})
\end{equation}
\citep{eadie71}. For a single polarization measurement ($n=1$) the
likelihood function becomes
\begin{equation}
%L_{p\phi}(p_0,\phi_0) & = & f(p,\phi), \\
L(P_0|P) \, \dif P_0 = \frac{1}{N_L} F(P|P_0) \dif P_0% \quad \mbox{and,} \\
%L_\phi(\phi_0,p_0) & = & G(\phi-\phi_0,p_0) ??.
\label{eq-likeli}
\end{equation}
where the normalization constant is
\begin{eqnarray}
%N_L & = & \int_{-\pi/2}^{\pi/2}\dif\phi_0\int_0^\infty\dif p\, L_{p\phi}(p_0,\phi_0) =
%\int_{-\pi/2}^{\pi/2}\dif\phi_0\int_0^\infty\dif p\, f(p,\phi) \nonumber \\ 
%& = &
N_L & = & 
\int_0^\infty F(P|P_0) \, \dif P_0 \nonumber \\
& = &
 P \sqrt{\frac{\pi}{2}}
\exp\left(-\frac{P^2}{4\sigma^2}\right) I_0\left(\frac{P^2}{4\sigma^2}\right).
%& = & \sqrt{\frac{\pi}{2}} \, p \, \me^{-p^2/4} \, I_0\left(\frac{p^2}{4}\right).
\label{eq-norm}
\end{eqnarray}
The likelihood function is plotted in Figure \ref{fig-rice2}.
% ----- FIGURE 2 -----

\subsection{Maximum Likelihood}

\begin{figure*}
\epsscale{1}
\plottwo{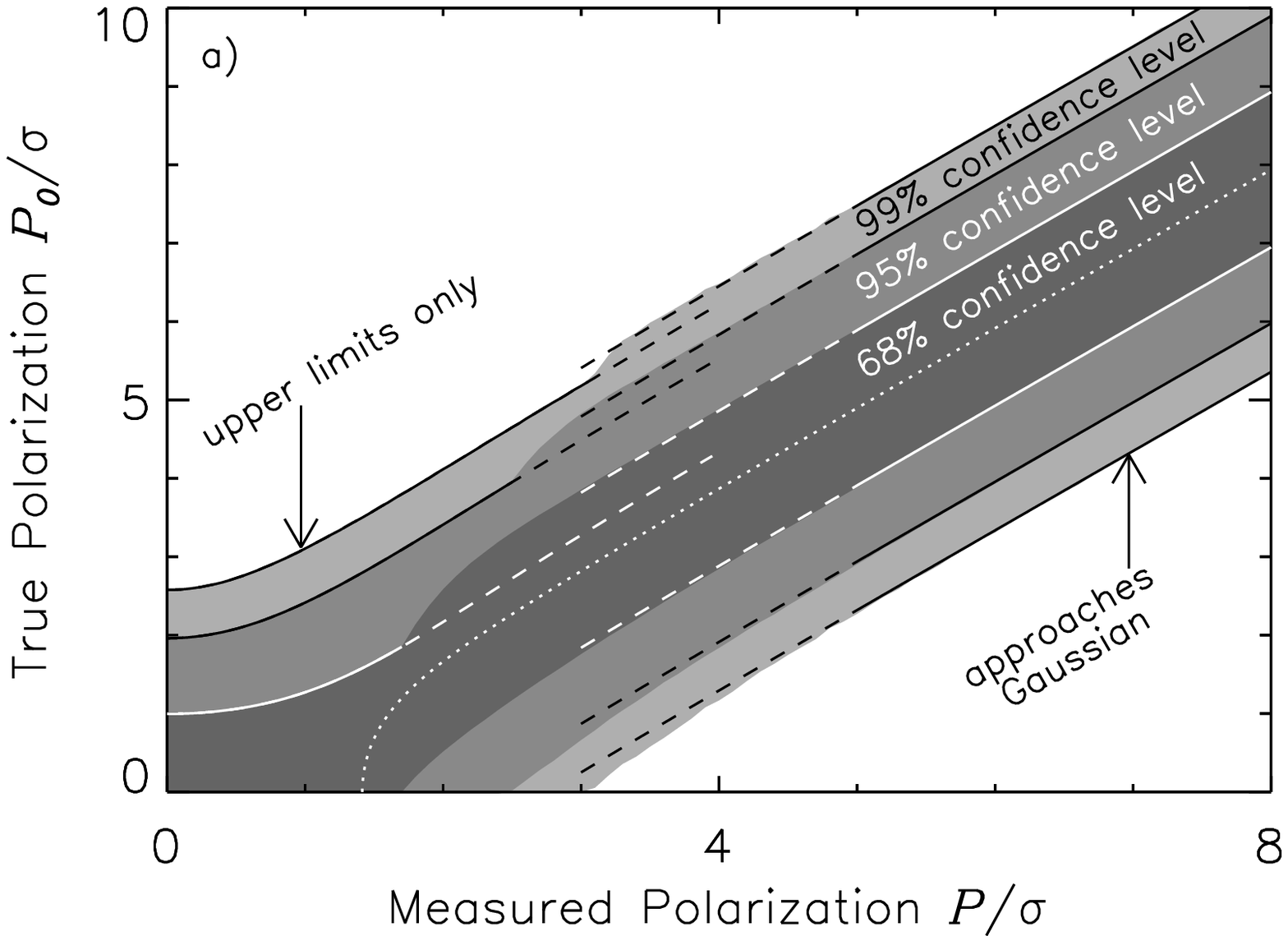}{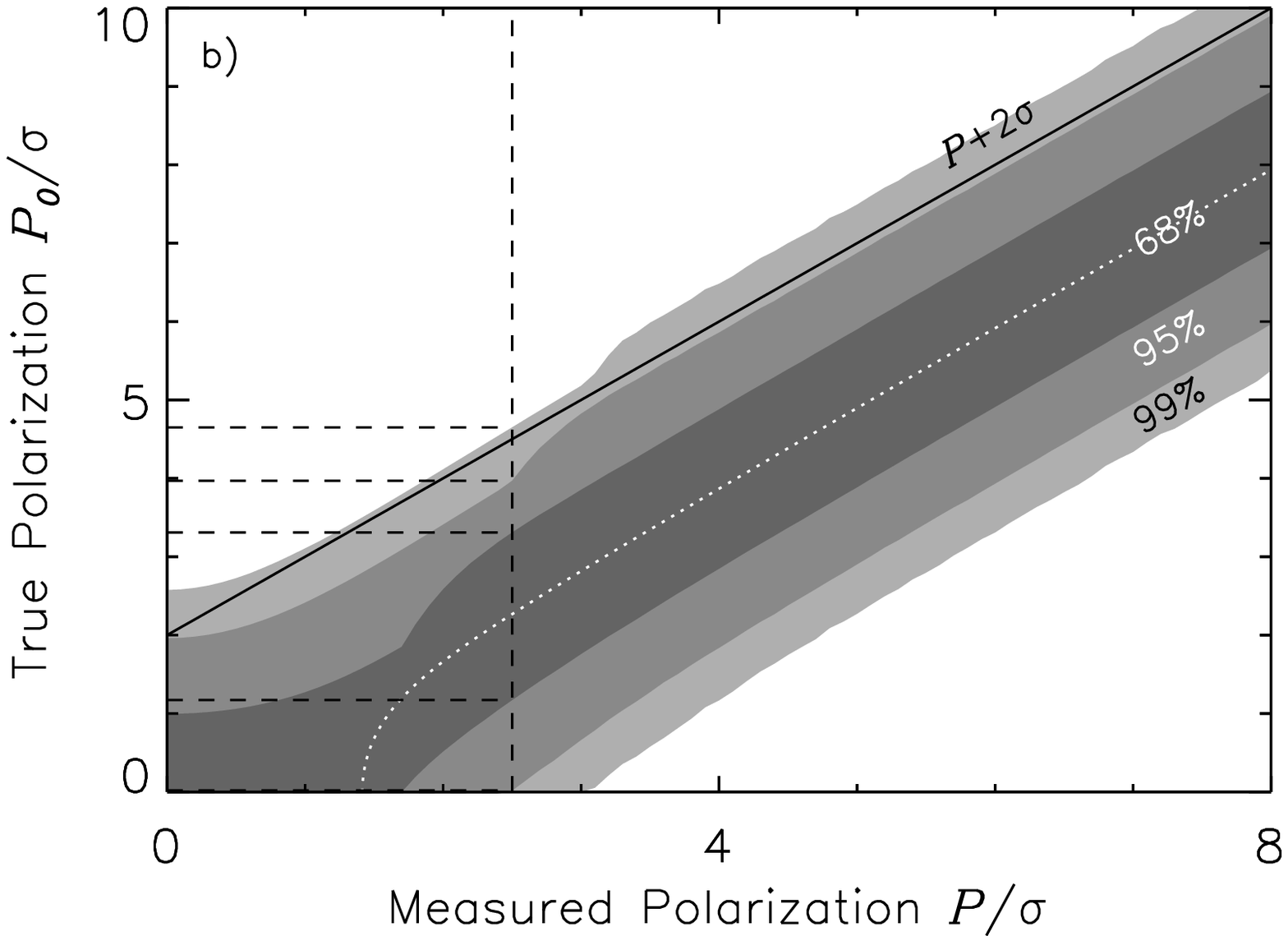}
%\plottwo{conflimit.eps}{example.eps}
\caption{Polarization confidence limits.  Grayscale indicates
  confidence regions of 68.0\%, 95.0\%, and 99.0\%. a) Solid curves
  labeled ``upper limits only'' and dashed line extrapolations
  indicate regions where the lower polarization limit $P_l =0$. At
  high signal-to-noise the confidence regions are bounded by those
  expected from a normal gaussian distribution (shown as labeled solid
  lines and dashed extrapolations). The dotted white curve indicates
  the maximum likelihood solution (eq.\ [\ref{eq-maxderiv}]). b) The
  solid line marks an upper limit which satisfies the condition $P_u =
  P+2\sigma$. Dotted lines illustrate the construction of confidence
  regions for a measured polarization of $P/\sigma=2.5$.
  \label{fig-conflimits}}
\end{figure*}

A ``best fit'' estimate of the true polarization can be extracted from
the measured polarization by utilizing the maximum likelihood.  Taking
the derivative of $L(P_0|P)$ we have
\begin{equation}
\left. \frac{\dif L}{\dif P_0}\right\vert_{\hat{P} =P_0} = 0 \; \Rightarrow \;
PI_1\left(\frac{P\hat{P}}{\sigma^2}\right) - \hat{P} I_0\left(\frac{P\hat{P}}{\sigma^2}\right) = 0
\label{eq-maxderiv}
\end{equation}
where $I_1()$ is the first order modified Bessel function and
$\hat{P}$ is the maximum likelihood estimate of $P_0$. Complete
solutions to this equation can be found numerically (Fig.\
\ref{fig-conflimits}a).  At the limits of low and high signal-to-noise
the solution is
\begin{eqnarray}
\hat{P} = 0 & \quad\mbox{for}\quad & P/\sigma < \sqrt{2} \label{eq-maxlik1}, \mbox{ and} \\
\hat{P} \approx \sqrt{P^2 - \sigma^2} & \quad\mbox{for}\quad & P/\sigma \gtrsim 3.
\label{eq-maxlik2}
\end{eqnarray}
SS85 have shown that the maximum likelihood estimator does not
completely correct for the positive bias of polarization measurements.
However, of the estimators investigated by SS85, the maximum
likelihood is superior for low signal-to-noise
($P_0/\sigma\lesssim0.7$).  For higher values, the most probable
estimator (that which maximizes the $F(P|P_0)$ in
Fig.\ \ref{fig-rice}) is best (see also \citealt{wardle74}).  For
$P_0/\sigma\gtrsim4$, both estimators approach that of equation
(\ref{eq-maxlik2}) (SS85).
% ----- FIGURE 3 -----

\subsection{Confidence Regions}

SS85 create confidence regions by integrating the probability density
function (eq.\ [\ref{eq-rice}]) over the measured polarization $P$.
To allow for the positive bias in the probability distribution they
construct confidence regions on $P_0$ by projecting the results onto
the $P_0$-axis. However, it is not clear that this approach completely
accounts for the asymmetry in the Rice distribution.

To determine the probability that a single measurement $P/\sigma$ is
drawn from a range of true polarizations $P_0/\sigma \in
[P_l/\sigma,P_u/\sigma]$ we integrate the likelihood function over the
parameter space of $P_0/\sigma$.  Following the approach of SS85,
limits of measure $\beta \times 100\%$ for each value of $P/\sigma$ can be
constructed from
\begin{eqnarray}
\int_0^{P_u/\sigma} L(P_0|P) \, \dif(P_0/\sigma) & = & \beta + \lambda
\quad \mbox{and}
\label{eq-upper} \\
\int_0^{P_l/\sigma} L(P_0|P) \, \dif(P_0/\sigma) & = & \lambda. \label{eq-lower}
\end{eqnarray}
Choosing $\lambda$ such that $(P_u - P_l)$ is minimized implies that
$L(P_l) = L(P_u)$ (for all $\lambda > 0$) but does not generate
symmetric confidence regions, especially for low $P/\sigma$.  From
Figure \ref{fig-conflimits}a we see that:

(1) For measured polarization values $P/\sigma < 1.7$, 2.4, and 3.0, the
lower end of the confidence region is bounded by zero for the 68.0\%,
95.0\%, and 99.0\% confidence levels, respectively.  It is not
possible to construct a truly bounded symmetric confidence region for
these polarizations; it is only possible to construct upper limits
($\lambda=0$).

(2) For polarizations $P/\sigma\gtrsim 4$ the confidence regions approach
those given by a normal gaussian distribution centered on the
de-biased value $\hat{P} \approx \sqrt{P^2-\sigma^2}$ ($68\%\rightarrow
1\sigma$, $95\%\rightarrow 2\sigma$, $99.0\%\rightarrow 2.6\sigma$).

Some authors have placed upper limits of ``high confidence'' on
polarization measurements using the criterion that $P+2\sigma< P_u$
(e.g.\ \citealt{archive}). While this criterion matches no
single confidence level, it does fall between the 95\% and 99\%
confidence levels even for low $P/\sigma$ (Fig.\ \ref{fig-conflimits}b).
For determining upper limits at high confidence this criterion seems a
reasonable (and perhaps more feasible) alternative to solving
equations (\ref{eq-upper}) and (\ref{eq-lower}).

Figure \ref{fig-conflimits}b illustrates the use of our results to
find confidence regions. To facilitate comparison of our results with
those of SS85, consider a polarization measurement of $P=2\%$,
$\sigma=0.8\%$.  Upper and lower confidence limits for $P/\sigma=2/0.8=2.5$
are given in Table \ref{tbl-compare}.  The differences between the two
sets of results increase with increasing confidence level.  These
differences arise because the integration over data space and
subsequent projection onto parameter space performed by SS85 is not
equivalent to our integration over parameter space.  However, these
two methods become congruent as the signal-to-noise increases and the
Rice distribution approaches a normal distribution.
% ----- TABLE 1 -----
\begin{deluxetable}{ccccc}[tb]
\tablecolumns{5}
\tablewidth{0pt}
\tablecaption{$P_0/\sigma$ confidence regions for $P/\sigma=2.5$ \label{tbl-compare}}
\tablehead{\colhead{Confidence Level} & \multicolumn{2}{c}{\underline{\citealt{simmons85}}} &
 \multicolumn{2}{c}{\underline{This Work}} \\
\colhead{(percent)}  & \colhead{Lower} & \colhead{Upper} & \colhead{Lower} & \colhead{Upper}}
\startdata
68.0\tablenotemark{a} & 1.4\phn & 3.3\phn & 1.2\phn & 3.3\phn \\
95.0 & 0.28 & 4.32 & 0.02 & 3.97 \\
99.0 & 0\phd\phn\phn & 4.96 & 0\phd\phn\phn & 4.65
\enddata
\tablenotetext{a}{\citet{simmons85} use a 67\% confidence level rather than 68\%.}
\end{deluxetable}

\section{Summary}

The probability distribution of polarization measurements is
asymmetric, resulting in a positive bias and complicating the
generation of uncertainties and confidence levels.  We generate
confidence intervals on the true polarization by integrating the
likelihood function over the parameter-space of the true polarization,
as opposed to the data-space of the measured polarization.  These
confidence levels are valid for both high and low signal-to-noise
measurements and approach those given by a normal distribution for
high signal-to-noise.

\acknowledgements{I thank Roger Hildebrand, Darren Dowell, Jessie
  Dotson, and Martin Houde for useful comments on drafts of this
  paper, and Tim Donaghy for introducing me to likelihood functions.}

%\bibliography{polar,hildebrand} \bibliographystyle{apj}

\end{document}